\documentclass[%
 aip,
 amsmath,amssymb,
 reprint,%
 citeautoscript,
]{revtex4-1}

\usepackage{graphicx}
\usepackage{dcolumn}
\usepackage{bm}

\usepackage[utf8]{inputenc}
\usepackage[T1]{fontenc}
\usepackage{mathptmx}
\usepackage{siunitx}
\usepackage{etoolbox}
\usepackage{tabularx}
\usepackage{amsmath,amsthm,amssymb}

\makeatletter                                                                             
\def\@email#1#2{%
 \endgroup                                                                                
 \patchcmd{\titleblock@produce}                                                           
  {\frontmatter@RRAPformat}                                                               
  {\frontmatter@RRAPformat{\produce@RRAP{*#1\href{mailto:#2}{#2}}}\frontmatter@RRAPformat}
  {}{}                                                                                    
}%
\makeatother                                                                              

\begin{document}

\preprint{AIP/123-QED}

\title{OrbNet Denali: A machine learning potential for biological and organic chemistry with semi-empirical cost and DFT accuracy}
\author{Anders S. Christensen}\thanks{Indicates equal contribution.}
\affiliation{Entos, Inc.\\ Los Angeles, CA 90027}
\author{Sai Krishna Sirumalla}\thanks{Indicates equal contribution.}
\affiliation{Entos, Inc.\\ Los Angeles, CA 90027}
\author{Zhuoran Qiao} 
\affiliation{Division of Chemistry and Chemical Engineering\\
California Institute of Technology\\
Pasadena, CA 91125}
\author{Michael B. O'Connor} \affiliation{Entos, Inc.\\ Los Angeles, CA 90027}
\author{Daniel G. A. Smith} \affiliation{Entos, Inc.\\ Los Angeles, CA 90027}
\author{Feizhi Ding} \affiliation{Entos, Inc.\\ Los Angeles, CA 90027}
\author{Peter J. Bygrave} \affiliation{Entos, Inc.\\ Los Angeles, CA 90027}
\author{Animashree Anandkumar}
\affiliation{Division of Engineering and Applied Sciences\\
California Institute of Technology\\
Pasadena, CA 91125}
\affiliation{NVIDIA\\ Santa Clara, CA 95051}
\author{Matthew Welborn} \affiliation{Entos, Inc.\\ Los Angeles, CA 90027}
\author{Frederick R. Manby} \affiliation{Entos, Inc.\\ Los Angeles, CA 90027}\email{fred@entos.ai}
\author{Thomas F. Miller III}\email{tom@entos.ai}
\affiliation{Entos, Inc.\\ Los Angeles, CA 90027}
\affiliation{Division of Chemistry and Chemical Engineering\\
California Institute of Technology\\
Pasadena, CA 91125}

\date{\today}

\begin{abstract}
We present OrbNet Denali, a machine learning model for electronic structure that is designed as a drop-in replacement for ground-state density functional theory (DFT) energy calculations.
The model is a message-passing neural network that uses symmetry-adapted atomic orbital features from a low-cost quantum calculation to predict the energy of a molecule.
OrbNet Denali is trained on a vast dataset of 2.3 million DFT calculations on molecules and geometries. This dataset covers the most common elements in bio- and organic chemistry (H, Li, B, C, N, O, F, Na, Mg, Si, P, S, Cl, K, Ca, Br, I) as well as charged molecules.
OrbNet Denali is demonstrated on several well-established benchmark datasets, and we find that it provides accuracy that is on par with modern DFT methods while offering a speedup of up to three orders of magnitude.
For the GMTKN55 benchmark set, OrbNet Denali achieves WTMAD-1 and WTMAD-2 scores of 7.19 and 9.84, on par with modern DFT functionals.
For several GMTKN55 subsets, which contain chemical problems that are not present in the training set, OrbNet Denali produces a mean absolute error comparable to those of DFT methods.
For the Hutchison conformers benchmark set, OrbNet Denali has a median correlation coefficient of $R^2$=0.90 compared to the reference DLPNO-CCSD(T) calculation, and $R^2$=0.97 compared to the method used to generate the training data ($\omega$B97X-D3/def2-TZVP),  exceeding the performance of any other method with a similar cost.
Similarly, the model reaches chemical accuracy for non-covalent interactions in the S66x10 dataset.
For torsional profiles, OrbNet Denali reproduces the torsion profiles of $\omega$B97X-D3/def2-TZVP with an average MAE of 0.12 kcal/mol for the potential energy surfaces of the diverse fragments in the TorsionNet500 dataset.

\end{abstract}

\maketitle

\section{Introduction}
Theoretical chemistry is based on the strategy of using approximate methods to make predictions for chemical problems.
Perhaps most notable within the field are approximations  to the time-independent Schrödinger equation, yielding the electronic energy that governs chemical potential energy surfaces.

While high-level wave function methods calculate energies to  high accuracy, these methods are often too slow for use in many areas of chemical research.
A pragmatic middle-ground is density functional theory (DFT) which can yield results close to chemical accuracy, often on the scale of minutes to hours of computational time.
In high-throughput applications, force-field (FF) and semi-empirical quantum mechanics (SEQM) methods are often used, providing results within fractions of a second of computational time, but at the cost of a loss in the reliability of the predictions.

Recent developments in machine learning models for chemistry have led to many new strategies for making energy predictions of molecules.\cite{Neuralnetworks_Scheffler2004,Neuralnetworks_BehlerParrinello2007,BehlerPerspective2016,BartokGAP2015,Huan2017,Zhenwei2015,Thompson2015,CovariantKernelsSandro2016,SmithANI2017,ANI2x,chmiela2017machine,Chmiela2019sGDML,mcgibbon2017improving,Grisafi2018,Glielmo2018,schutt2018schnet,schutt2019schnetpack,Comorant2019,DimeNet,unke2019physnet,mobml1,mobml2,Faber2018,christensen2019operator,christensen2020fchl,deephf,qiao2020orbnet,qiao2020multi,DeepMDZhang2018}
Relatively few of these, however, have sufficient accuracy and breadth of training to provide a viable replacement for quantum chemistry methods, such as DFT, without needing to retrain the model for specialized purposes.
The difficulty of constructing such machine learning models that can make reliable predictions across a large fraction of chemical space is two-fold: (i) the model must capture the underlying physics, and (ii) relevant and well-curated training data covering the relevant chemical problems must be available.

The current paper introduces OrbNet Denali to address these issues.
OrbNet Denali is a scale-up of the previously introduced orbital-based neural network potential, OrbNet, both in terms of model  size and training data.\cite{qiao2020orbnet,qiao2020multi}
On the technical side, this scale-up includes a number of practical and architectural engineering improvements to the OrbNet code which, in turn, enables training on an expanded dataset with much a wider coverage of chemical space.
OrbNet is a message passing neural network\cite{gilmer2017neural} that encodes a molecular system in graphs based on features from a low-cost quantum calculation.
In this work specifically, the low-cost quantum calculation is based on the GFN1-xTB method,\cite{gfn1} and the features are generated from the Fock, density, orbital centroid distance, core Hamiltonian, and overlap matrices evaluated in the symmetry-adapted atomic orbital (SAAO) basis.\cite{qiao2020orbnet}
Graph-node features correspond to diagonal SAAO matrix elements, and edge features correspond to off-diagonal SAAO matrix elements.
%
%

%
OrbNet Denali is trained on a new dataset of 2.3 million
molecules generated from a variety of sources.
The combined dataset covers a range of organic molecules and chemistries, with protonation and tautomeric states, non-covalent interactions, common salts, and counterions, spanning the most common elements in bio- and organic chemistry (H, Li, B, C, N, O, F, Na, Mg, Si, P, S, Cl, K, Ca, Br, I). 
These combined developments enable reliable energy prediction with an accuracy comparable to DFT, i.e., the same method used to calculate the energies corresponding to the geometries in the training data.
We demonstrate the performance and transferability of the trained model by benchmarking OrbNet Denali across the many diverse chemical problems in the well-established collection of benchmark sets, GMTKN55.\cite{GMTKN55}
Furthermore, we demonstrate the performance of OrbNet Denali on several other essential tasks, such as conformer scoring, non-covalent interactions, and torsional profiles.\cite{Hutch,S66Rezac2011,TorsionNet500}

\section{Methods}
\subsection{Training data}
\label{sec:training_data}

The main advance of this work is the expansion of the OrbNet Denali training set.
The collection of training data is as described below.

\subsubsection{ChEMBL molecule conformers}
The ChEMBL27 database\cite{chembl27} was downloaded from the ChEMBL web service.\cite{ChemblDavies2015,ChemblMendez2018}
All SMILES strings containing 50 or fewer atoms of the elements C, O, N, F, S, Cl, Br, I, P, Si, B, Na, K, Li, Ca, or Mg and no isotope specifications were kept.
SMILES strings that did not resolve to a closed-shell Lewis structure were discarded. All SMILES strings corresponding to molecules in the Hutchison conformer benchmark set \cite{Hutch}  were removed from our training dataset. (In  Sec. \ref{sec:hutchison}, these molecules will be used to validate the model.)

From this subset, a final surviving selection of 116,943 unique SMILES strings corresponding to neutral molecules was randomly chosen.
Up to four conformers for each SMILES string were initially generated through the \textsc{Entos Breeze} version 0.1.5 conformer generator and optimized at the GFN1-xTB level.\cite{gfn1}
For each of these four energy-minimized conformers, non-equilibrium geometries were generated using \textsc{Entos Breeze} through either normal-mode sampling\cite{SmithANI2017} at 300K or \textit{ab initio} molecular dynamics (AIMD) sampling for 200fs at 500K; in both cases at the GFN1-xTB level of theory.
These thermalization methods were selected randomly for each molecule with equal weight.
This procedure resulted in a total of 1,771,191 equilibrium and non-equilibrium geometries.

\subsubsection{Protonation states and tautomers}
A subset of 26,186 SMILES strings were randomly chosen from the list of filtered ChEMBL SMILES strings described in the above text.
For each of these, up to 128 unique protonation states were identified using \textsc{Dimorphite-DL}\cite{DimorphiteDL} version 1.2.4 and four of these protonation states were selected at random.
The same conformer generation algorithm and non-equilibrium geometry sampling algorithms as described above were applied to the four protonation states, resulting in a total of 215,866 unique geometries.
Additionally, the protonation-state sampling was applied to the SMILES strings of the molecules in the QM7b dataset, resulting in 9,413 unique molecule graphs and 30,622 geometries.\cite{Blum,VonLilienfeld2013}

\subsubsection{Salt complexes and non-bonded interactions}
From the list of filtered ChEMBL SMILES strings, a number of SMILES strings were selected and randomly paired with between one to three salt molecules from the list of common salts in the ChEMBL Structure Pipeline.\cite{ChemblPipelineBento2020}
This procedure resulted in a total of 21,735 salt complexes.
For each of these complexes, four conformers were created through our conformer pipeline, and normal-mode sampling as described above was used to generate four non-equilibrium geometries for each conformer.
This resulted in 271,084 unique geometries.
Additionally, the structures in the JSCH-2005\cite{JSCH} and the sidechain-sidechain interaction (SSI) subset of the BioFragment Database\cite{BFGDBBurns2017} (BFDb) were added to the dataset.

\subsubsection{Small molecules}
To avoid biasing the datasets to represent only large drug-like molecules, a list of common chemical moieties and bonding patterns in organic molecules was created, and combined to generate small molecules with up to 21 heavy atoms and relatively “exotic” compositions, resulting in around 15,000 SMILES strings.
For each of these, SMILES strings were created by randomly substituting hydrogen atoms for halogens, and carbon for silicon.
This procedure resulted in a total of 40,565 SMILES strings, for which conformers were generated through our conformers pipeline, resulting in a total of 94,588 unique geometries.

\subsection{QM calculations}
All DFT single-point calculations for the OrbNet Denali training set were carried out in \textsc{Entos Qcore} version 0.8.17 at the $\omega$B97X-D3/def2-TZVP level of theory using in-core density fitting with the \texttt{neese=4} DFT integration grid.\cite{wb97xd,Weigend2005,entos,Polly2004,Neese2011}

In order to gauge the performance of OrbNet Denali, a number of additional reference calculations on the various test sets presented in Section~\ref{sec:results} were carried out using a number of low-cost models.
B97-3c calculations were carried out in \textsc{Entos Qcore},\cite{Brandenburg2018,entos} while reference GFN$n$-xTB calculations were carried out in the \textsc{xTB} program.\cite{GFN1xTBGrimme2017,GFN2xTBBannwarth2019,XTBBannwarth2020}
ANI-1ccx and ANI-2x calculations were carried out using the \textsc{TorchANI} implementation\cite{smiti_transfer_2018,ANI2x,TorchANI2020} and, lastly, MMFF94 calculations were carried out using the Open Babel command-line tool.\cite{MMFF94,OpenBabel}

\subsection{Training details}
\begin{table*}
\caption{\label{tab:model_comparison} Comparison of models reported in this work to previously reported OrbNet models.}
\begin{tabular}{lrrrrr}
\hline
Model                  & \# trainable parameters & \# training data & Description of training data                                               & Multitask? &  \\ \hline
Qiao et al.\cite{qiao2020orbnet}            & $\sim$1.8 million                & $\sim$ 0.2 million & QM7b-T, QM9, GDB-13-T, DrugBank-T                           & No         &  \\
Qiao et al.\cite{qiao2020multi}            & $\sim$1.8 million                & $\sim$ 0.2 million & QM7b-T, QM9, GDB-13-T, DrugBank-T                           & Yes        &  \\
``Denali (10\%)'' (this work) & $\sim$21 million          & $\sim$ 0.2 million & 10\% of data described in Sec. \ref{sec:training_data} & No         &  \\
``Denali'' (this work)  & $\sim$21 million                & $\sim$ 2.3 million & All data described in Sec. \ref{sec:training_data}          & No         &  \\ \hline
\end{tabular}
\end{table*}

\textsc{PyTorch} \cite{PYTORCH_NEURIPS} v1.7.1 and the Deep Graph Library (\textsc{DGL}) \cite{DGL} v0.6 were used to implement and train OrbNet Denali. 
PyTorch's Distributed Data Parallel (DDP) \cite{li2020pytorch} strategy was used to train the model on multiple GPUs using data parallelism. 
OrbNet Denali was trained on the OLCF Summit supercomputer using 96 NVIDIA V100-SXM2 (32G) GPUs with a batch size of 4 per GPU for 300 epochs, totaling 6912 GPU-hours of training. 
The learning rate was linearly warmed-up for the first 100 epochs and cosine annealed \cite{cosine_annealing} to zero for the remaining 200 epochs. 
During this process, the maximum learning rate was $3\times 10^{-4}$ and the model parameters were optimized with the Adam optimizer.\cite{kingma2014adam}
The training dataset of size 1.8\;TB was randomly split into four shards of equal sizes.
Each Summit node, comprising 6 GPUs, was assigned to one of these four shards such that each shard was used on 1/4 of the nodes allocated for the training job.
For example, training with 16 nodes results in each shard assigned to 4 nodes.

An additional model (``OrbNet Denali 10\%'') was trained on a 10\% subset of the training data, constructed by randomly sampling 10\% of molecules or complexes in each of the various subsets of the full training dataset.
All other training details were the same.
Finally, we use a small set of randomly selected geometries from each subset described in Section \ref{sec:training_data} to validate and track the model convergence during training. 
Table \ref{tab:model_comparison} provides a comparison of models presented in this work to previously reported OrbNet models.

The OrbNet Denali model and feature generation employed in this work are based on the model previously reported in Ref. \citenum{qiao2020multi}. 
Compared to that work, four changes were made.
First, the attention mechanism was replaced with the FAVOR+ mechanism \cite{performer_attention}.
This resulted in greatly decreased memory usage and negligible test accuracy degradation. 
Second, the number of message passing steps was increased from two to three. 
Third, batch normalization layers\cite{ioffe2015batch} were replaced with layer normalization layers\cite{ba2016layer} without affine transformations to facilitate multi-GPU training.
Finally, the regression labels were modified to account for charged molecules. The regression labels are 
\begin{equation}
    E^{\textrm{ML}} \approx E^{\textrm{DFT}} - E^{\textrm{GFN1}} - \Delta E^{\textrm{fit}}_\textrm{atoms},
\end{equation}
where $E^\textrm{DFT}$ is the reference DFT (i.e. $\omega$B97X-D3/def2-TZVP) energy and $E^\textrm{GFN1}$ is the GFN1-xTB energy.\cite{ramakrishnan2015big}
In this work, $\Delta E^\textrm{fit}_\textrm{atoms}$ is given by
\begin{equation}
    \Delta E^{\textrm{fit}}_\textrm{atoms} = \sum_i \Delta E_{Z_i}^\textrm{fit} - \Delta E_{e^-}^\textrm{fit} q,
\end{equation}
where $i$ indexes atoms within a molecule, $Z_i$ is the atomic number of atom $i$, and $q$ is the total charge of the molecule.
$\Delta E_{Z_i}^\textrm{fit}$ and $\Delta E_{e^-}^\textrm{fit}$ are parameters and are fit to $E^{\textrm{DFT}} - E^{\textrm{GFN1}}$ with ordinary least squares prior to training.

The loss function used here is similar to that of Ref.~\citenum{qiao2021unite}, and contains two terms.
The first term minimizes the error on the absolute energy of the every sample in the training set, while the second term relates to the relative energies of different geometries of the same molecule:
\begin{eqnarray}
\mathcal{L} & = & \mathcal{L}_2\left( \hat{E}\left( \eta, b_\eta \right), E \left( \eta, b_\eta \right) \right) \\
&& + \beta \mathcal{L}_2\left( \hat{E}\left( \eta, b_\eta \right) - \hat{E}\left( \eta, \hat{b}_\eta \right), E \left( \eta, \hat{b}_\eta \right) - E\left( \eta, \hat{b}_\eta \right) \right)\nonumber
\end{eqnarray}
Here, $\mathcal{L}_2$ denotes the L2-loss function $\mathcal{L}_2\left( \hat{y}, y \right) = \| \hat{y} - y \|^2_2$, while $b_\eta$ is the geometry of a molecule sampled from the pool of geometries $\{ b_\eta \}$ for each molecule, $\eta$, within each minibatch, and $\hat{b}_\eta$ denotes a different randomly sampled geometry from the same molecule, $\eta$.
We used $\beta=9$ in this work. 

\section{Results}
\label{sec:results}
\subsection{The GMTKN55 collection}
    \begin{figure}
        \centering
        \includegraphics[width=\linewidth]{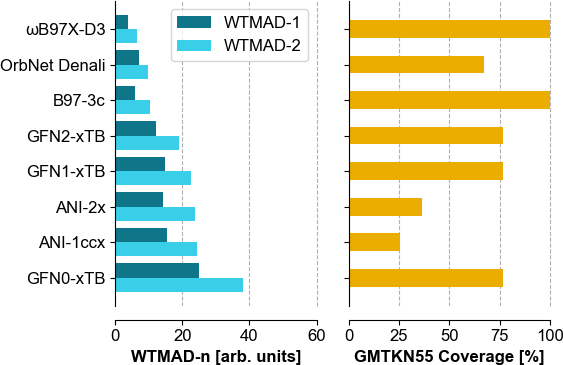}
        \caption{\label{fig:gmtkn55_coverage} Statistics over the accuracy and coverage of the GMTKN55 dataset\cite{GMTKN55} is shown for a selection of methods, sorted by WTMAD-2 scores, relative to reference high-level\cite{GMTKN55} estimates.
        The left panel shows the aggregated WTMAD-1 and WTMAD-2 metrics in arbitrary units, calculated on only those subsets that are covered by each method.
        The right panels shows the percentage of GMTKN55 subsets consisting only of molecules with elements, charged states, and spin states that are allowed within each model.
        The def2-TZVP basis set\cite{Weigend2005} was used for the $\omega$B97X-D3 calculations shown in this figure.
            }
    \end{figure}
The General Main-group Thermochemistry, Kinetics, and Non-covalent Interactions 55 (GMTKN55) dataset\cite{GMTKN55} is a collection of 55 datasets aimed at probing the accuracy of quantum mechanical (QM) methods across a variety of chemical problems, ranging from reaction energies and electronic properties to non-covalent interaction energies and conformational properties.
The dataset consists of 55 individual subsets with a total of 1505 relative energies based on 2462 single-point calculations.
The high-level reference energies for the molecules in GMTKN55 are best-estimates calculated using a range of extrapolative protocols based on CCSD and CCSD(T) calculations collected from several different sources, and most recently updated in Ref.~\citenum{GMTKN55}.

Commonly, the performance of QM methods on GMTKN55 is presented via aggregated scores based on weighting of the mean absolute deviation to a reference, the WTMAD-1 or WTMAD-2 scores, with the difference between the two being the relative weighting of the individual subsets.\cite{GMTKN55}

We note that we are comparing QM models and ML models that do not always provide predictions for every subset of GMTKN55, for example due to the lack of capabilities to handle certain elements, charge- or spin-states in a model.
In order to enable a fair comparison between models with varying coverage, we present WTMAD scores that are calculated only over the GMTKN55 subsets that contain elements, charge- and spin-states that are supported for each model, and set the weight to 0 for the MAD for unsupported subsets.
For the GFN$n$-xTB methods, we also exclude those dataset that contain spin-states other than singlets.
Although we do note that while it is possible to run xTB calculations with unpaired electrons, these will not yield, for example, singlet-triplet splittings.\cite{XTBBannwarth2020}
An overview of the supported GMTKN55 subsets of each ML and xTB model compared in this text is given in the Supplementary Information, Table~\ref{tab:si_gmtkn55_coverage}.
%

%
%

Calculating the WTMAD-1 and WTMAD-2 scores for OrbNet Denali as described above, we find the WTMAD-1 and WTMAD-2 scores to be 7.19 and 9.84 against the high-level reference energies, respectively.
We note that a number of the subsets included in these weighted metrics contain examples of chemistry that are not present in the dataset used to train OrbNet Denali; for example water clusters, C$_{60}$ molecules, transition states geometries, etc.
If these "extrapolative" subsets (PA26, NBPRC, ALK8, C60ISO, WCPT18, PX13, INV24, BHROT27, BHPERI, BHDIV10, WATER27 and IDISP) are removed from the pool, the WTMAD-1 and WTMAD-2 scores are slightly changed by $-0.60$ and $+0.19$ to 6.59 and 10.03, respectively, which indicates that OrbNet Denali is able to yield reasonable predictions for many chemical problems that are necessarily not covered by the training set.

When the weighted scores are calculated against $\omega$B97X-D3/def2-TZVP reference energies (the same method used to generate the OrbNet Denali training data), the WTMAD-1 and WTMAD-2 scores are 6.42 and 8.40, respectively.
For the version of OrbNet Denali trained on 10\% of the data, the WTMAD-1 and WTMAD-2 scores are 7.40 and 11.41, respectively, against the $\omega$B97X-D3/def2-TZVP reference, demonstrating positive effects of increasing the dataset size.

We note that the WTMAD-1 and WTMAD-2 between $\omega$B97X-D3/def2-TZVP and the high-level reference energies are 3.37 and 5.87, respectively, which in some sense constitutes an upper bound for the accuracy versus high-level reference energies of an OrbNet model trained on $\omega$B97X-D3/def2-TZVP data.

A graphical overview of the performance of the OrbNet Denali, and the models discussed in the following, compared to $\omega$B97X-D3/def2-TZVP reference data is displayed in Fig.~\ref{fig:gmtkn55_performance} and in the Supplementary Information Fig.~\ref{fig:si_gmtkn55_extrapolative}.

In comparison, the popular semi-low-cost DFT method B97-3c\cite{Brandenburg2018} has WTMAD-1 and WTMAD-2 values for GMTKN55 of 5.76 and 10.22, respectively, compared to the high-level references, very close to the OrbNet Denali scores.
For this dataset, OrbNet Denali is roughly 100 times faster than B97-3c.

Another popular family of low-cost QM methods is the series of GFN$n$-xTB ($n\in \{0, 1, 2 \}$) methods.\cite{GFN1xTBGrimme2017,GFN2xTBBannwarth2019,XTBBannwarth2020}
%
For these methods we find the WTMAD-1 values to be 24.8,  14.7, and 12.0  for GFN0-xTB,\cite{gfn0}  GFN1-xTB,\cite{gfn1} and GFN2-xTB\cite{gfn2} respectively, with the same series of WTMAD-2 numbers being 38.0, 22.6, and 18.9.
We point out that GFN1-xTB is the baseline method used to generate the input for OrbNet Denali, and that OrbNet Denali yields a substantial improvement to this level of theory, resulting in DFT-quality energy predictions across the covered parts of GMTKN55.

For the machine learning potentials ANI-1ccx and ANI-2x, we calculate the WTMAD-$n$ scores over the subsets that only contain neutral singlet molecules with the elements that are covered by the individual methods (see Supplementary Information, Table~\ref{tab:si_gmtkn55_coverage}).\cite{Smith2017,ANI2x,smiti_transfer_2018,Smith2020}
For the ANI model that is parametrized against CCSD(T) reference data, namely ANI-1ccx,\cite{smiti_transfer_2018} we find the WTMAD-1 and WTMAD-2 values to be 15.5 and 24.2, respectively compared to CCSD(T) reference data.
For the ANI-2x model\cite{ANI2x} which, similarly to OrbNet Denali, is parametrized on DFT-level data, the WTMAD-1 and WTMAD-2 are 14.2 and 23.9, respectively, with respect to CCSD(T) reference data.
These WTMAD numbers are also found in Fig.~\ref{fig:gmtkn55_coverage}.
A graphical overview of the performance of ANI-2x can be found in Fig.~\ref{fig:gmtkn55_coverage} compared to $\omega$B97X-D3/def2-TZVP reference data.
We note that ANI-2x is parametrized from data calculated at the $\omega$B97X/6-31G(d) level and therefore the comparison in Fig.~\ref{fig:gmtkn55_coverage} does not reflect the true performance of ANI-2x compared to its underlying method, since differences between the two DFT methods are expected.

In terms of coverage of common chemistry problems to which a general-purpose machine learning potential can be applied, OrbNet Denali has the broadest coverage of GMTKN55, to our knowledge, of any currently existing ML-based potential.
OrbNet Denali covers 37 out of the 55 subsets; the OrbNet Denali training set does not cover the elements He, Be, and Al as well as a few heavy metals, nor spin states other than singlets, for example used to calculate ionization potentials and electron affinities.
We note that even when extrapolating out of the training distribution to these other subsets, OrbNet Denali provides reasonable, but less accurate, results due to its baseline method, GFN1-xTB.
The corresponding numbers for covered GMTKN55 subsets for ANI-1ccx and ANI-2x are 14 and 20, respectively.
ANI-1ccx covers only neutral, singlet molecules with the elements H, C, N, and O, while ANI-2x extends this coverage to the elements F, Cl, and S.\cite{ANI2x,Smith2020}
The family of GFN$n$-xTB methods has been parametrized on data containing elements up to Rn ($Z=86$), but handles systems with odd numbers of electrons in a spin-averaged way, and therefore covers 42 of the 55 GMTKN subsets.
The basis for these numbers is also summarized in the Supplementary Information, Table~\ref{tab:si_gmtkn55_coverage}. 

A graphical overview of WTMAD-$n$ values and the coverage of GMTKN55 subsets for each method can be found in Fig.~\ref{fig:gmtkn55_coverage}.

\begin{figure*}
    \centering
    \includegraphics[width=0.6\linewidth]{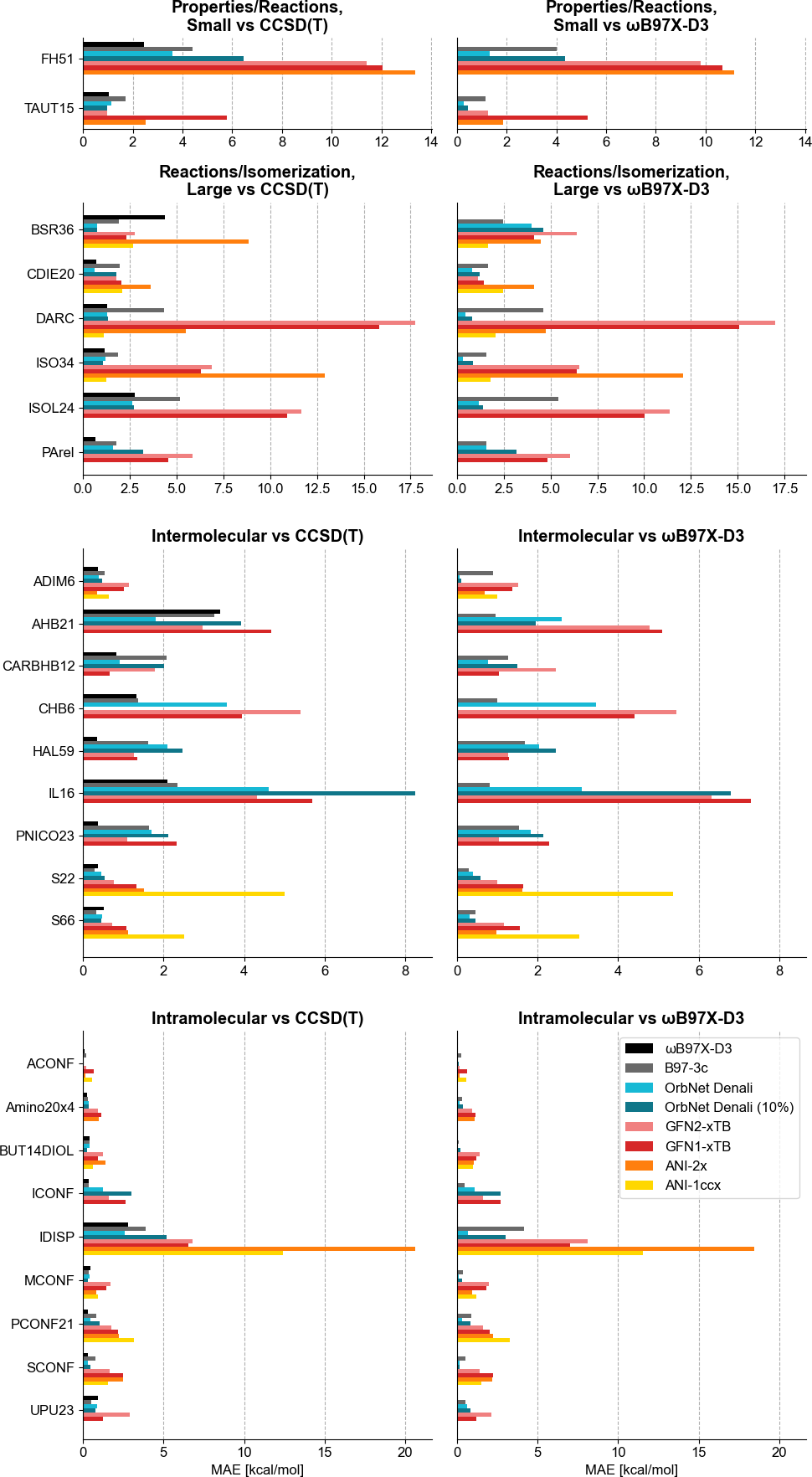}
    \caption{The figure shows the performance of OrbNet Denali on the GMTKN55 subsets that only contain elements, charge- and spin-states that are supported by OrbNet Denali, and where molecules with similar chemistry are found in the training set for OrbNet Denali.  In the left column, the MAE values are with respect to the CCSD(T) references in the GMTKN55 dataset\cite{GMTKN55} while, in the right column, the MAE values are with respect to $\omega$B97X-D3/def2-TZVP reference energies (the same method used to generate the OrbNet training data).\label{fig:si_gmtkn55_interpolative}}
\end{figure*}

A full overview of the performance of every method mentioned in this section compared to both CCSD(T) and $\omega$B97X-D3/def2-TZVP reference values for all the 37 relevant GMTKN55 subsets is presented in the Supporting Information, Figures~\ref{fig:si_gmtkn55_extrapolative} and~\ref{fig:si_gmtkn55_interpolative}.

\begin{figure}
    \centering
    \includegraphics[width=\linewidth]{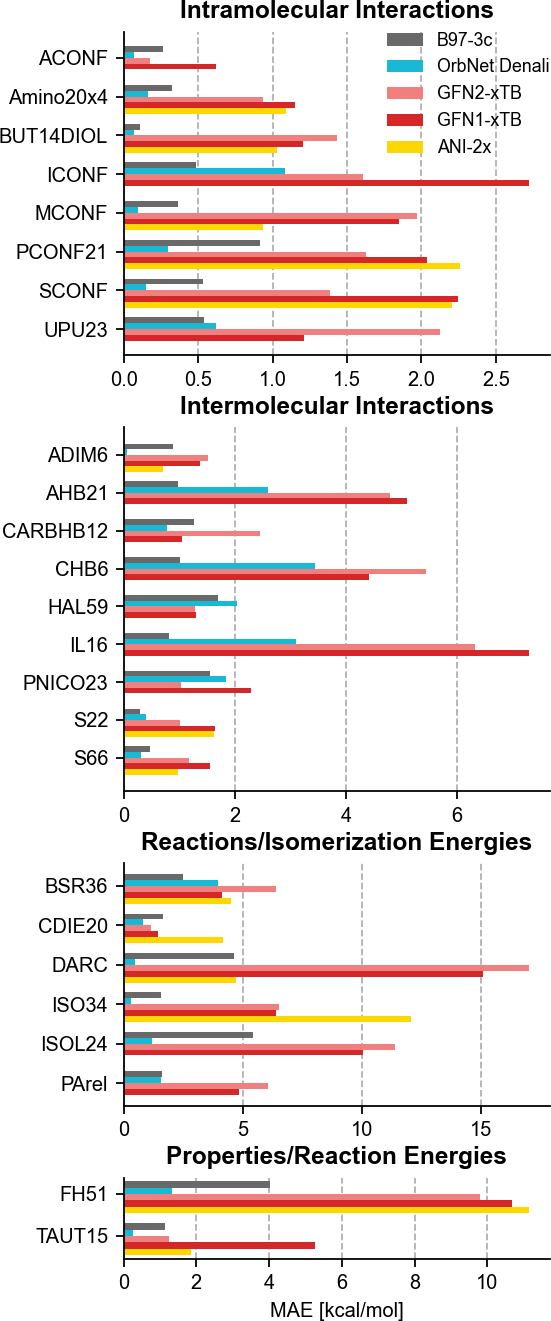}
    \caption{\label{fig:gmtkn55_performance} The chart shows the MAE in kcal/mol for the subsets of the GMTKN55 that are covered by the OrbNet Denali training data relative to $\omega$B97X-D3/def2-TZVP.
    For ANI-2x, values are left out for those subsets that contain elements or charge states that are not allowed by ANI-2x.
    }
\end{figure}

\subsection{Conformer scoring}
\label{sec:hutchison}

Accurate determination of the ensemble of thermally accessible conformers is key to modelling molecules.
In this section, we present results for a recent benchmark of conformer energetics.\cite{Hutch}
This benchmark contains up to ten poses for each of $\sim$700 drug-like molecules. Each molecule is comprised of C, H, N, O, S, Cl, F, P, Br, I, and contains between nine and fifty heavy atoms with a total charge between -1 and +2. 

Accuracy in this benchmark for a given method is reported as the median $R^2$ value and is determined as follows.
For every molecule, the correlation coefficient ($R^2$) is computed between the conformer energies of that molecule and the reference DLPNO-CCSD(T) energies.
The median is then taken over the set of $R^2$ values corresponding to all molecules in the benchmark set.

\begin{figure}
    \centering
    \includegraphics[width=\linewidth]{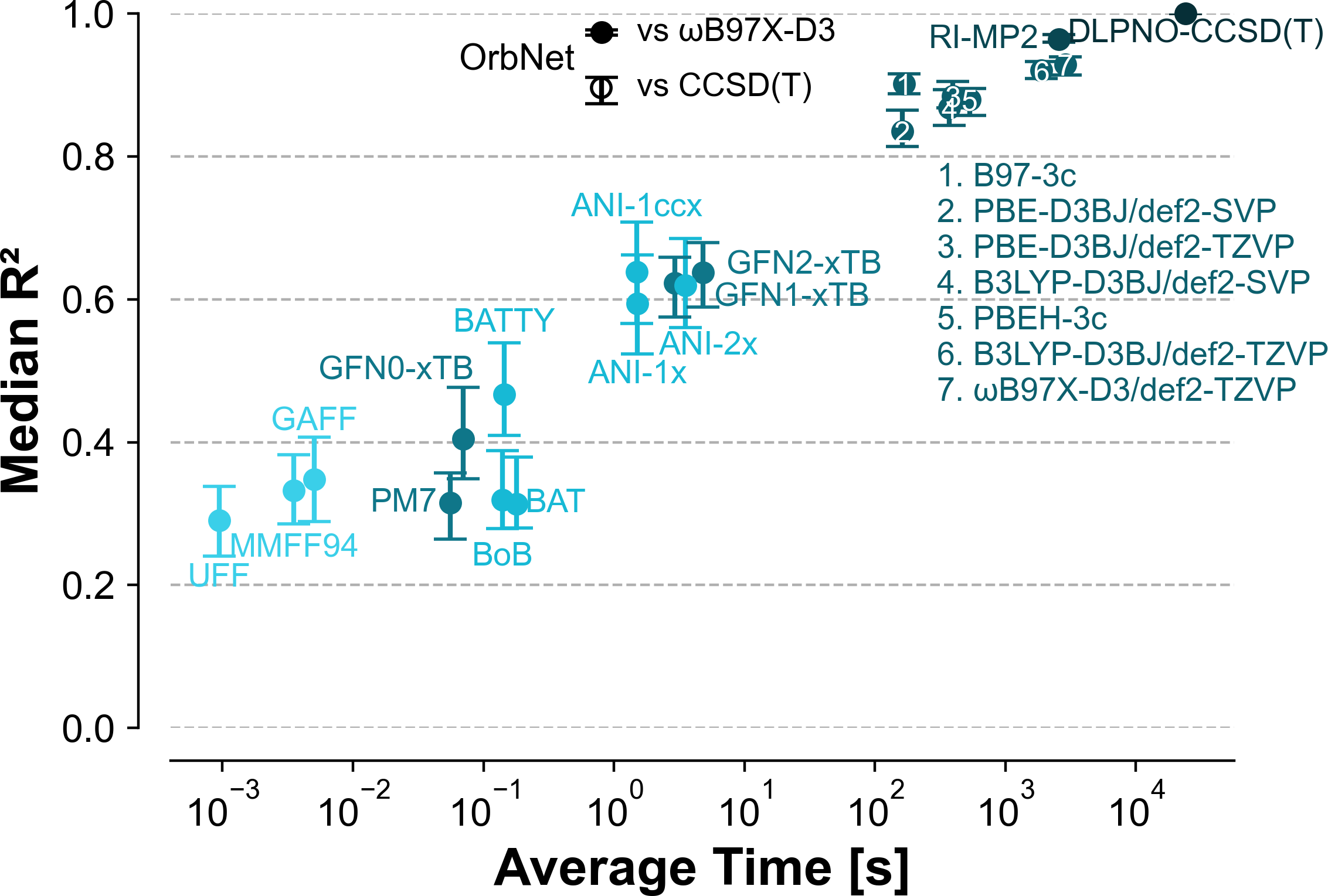}
    \caption{\label{fig:hutchison_conformers} A comparison between computational cost and the resulting accuracy for a number of methods for the Hutchison conformer benchmark set.
    The horizontal axis shows the average time of a single conformer energy computation, while the vertical axis denotes the median $R^{2}$ correlation coefficient for the molecules in the dataset.
    Error bars denote the 95\% confidence interval for this number and are obtained through bootstrapping.
    The median correlation coefficient versus DLPNO-CCSD(T) is shown for all methods (filled circles).
    Additionally, the median correlation coefficient versus $\omega$B97X-D3/def2-TZVP reference energies is shown for OrbNet Denali (open circle).
    This reference corresponds to the level of theory used to train the model.
    All data in this figure come from Ref. \citenum{Hutch}, with the exception of the two OrbNet Denali data points.
    For timings, all methods are run on a CPU, following the procedure described in Ref. \citenum{qiao2020orbnet}.
        }
\end{figure}

Fig.~\ref{fig:hutchison_conformers} provides a direct comparison of OrbNet Denali to a representative sample of computational chemistry methods, including force fields, machine learning, semi-empirical, density functional theory, and wavefunction theory.\cite{Hutch,qiao2020orbnet}
For methods other than OrbNet Denali, a strong correlation is observed between accuracy and the logarithm of the average execution time of the method.
OrbNet Denali breaks this trend, providing a median $R^2$ of 0.90 $\pm$ 0.02 versus the reference DLPNO-CCSD(T) at an average execution time of approximately one second per molecule.
Here, the uncertainty refers to the 95\% confidence interval and is obtained by bootstrapping the dataset.
GFN1-xTB, the method used to generate input for OrbNet Denali, provides a median $R^2$ of 0.62 $\pm$ 0.04 with a similar execution time to OrbNet Denali.
The median $R^2$ between OrbNet Denali and $\omega$B97X-D3/def2-TZVP, the same method used to generate the training data, is 0.973 $\pm$ 0.004, highlighting that OrbNet Denali is able to learn the its underlying method to high accuracy.
The OrbNet model train on 10\% of the data reaches a lower median $R^2$ value 0.94 $\pm$ 0.01 compared to $\omega$B97X-D3/def2-TZVP, demonstrating the positive effects of increasing the amount of training data.
Compared to the $\omega$B97X-D3/def2-TZVP level of theory which provides a similar accuracy versus DLPNO-CCSD(T) with a median $R^2$ of 0.92 $\pm$ 0.02, OrbNet Denali results in a more than thousandfold speedup.
This number also serves as an upper bound for the accuracy of a model trained on $\omega$B97X-D3/def2-TZVP data, and suggests that to increase the median $R^2$ for OrbNet Denali compared to DLPNO-CCSD(T), it is necessary to train on data that exceeds the accuracy of DFT.

\subsection{Non-covalent interactions (S66x10)}
A standard benchmark for the accuracy of non-covalent interaction is the S66x10 benchmark set.\cite{S66Rezac2011,S66DiLabio2013,Smith2016}
This dataset consists of 66 different molecular dimers and their equilibrium geometries, along with 9 additional displacements along the center-of-mass axis and corresponding CCSD(T)/CBS extrapolated binding energies.

For OrbNet Denali, we find the MAE and RMSE to CCSD(T)/CBS to be 0.75 and 1.01 kcal/mol respectively.
We note that these numbers are very close the the MAE and RMSE for the method used to generate the training data, $\omega$B97X-D3/def2-TZVP, at 0.70 and 0.91 kcal/mol.
Comparing OrbNet Denali to $\omega$B97X-D3/def2-TZVP we find smaller MAE and RMSE values at 0.46 and 0.65 kcal/mol, respectively.
For the OrbNet model trained on only 10\% of the data, these numbers increase to 0.67 and 0.85 kcal/mol compared to $\omega$B97X-D3/def2-TZVP, respectively, suggesting that the increased training data size is indeed beneficial, but also that it is impossible for the model to substantially surpass the accuracy of the training data.
The numbers referred to in this section are summarized in Table \ref{tab:s66x10}.

\begin{table}
\caption{
    \label{tab:s66x10}The MAE and RMSE binding energies for the S66x10 benchmark versus CCSD(T)/CBS reference binding energies.\cite{S66Rezac2011,S66DiLabio2013} 
    For rows marked with an asterisk ($^{*}$), OrbNet Denali predictions are compared to binding energies calculated at the the $\omega$B97X-D3/def2-TZVP level.
    The latter reference corresponds to the same method used to generate the training data for OrbNet Denali. 
}
\begin{tabular}{lrr}
\hline
    Model                       & MAE              & RMSE   \\
                                & [kcal/mol]        & [kcal/mol] \\\hline
    OrbNet Denali               &  0.75             &  1.01 \\
    $\omega$B97X-D3/def2-TZVP   &  0.70             &  0.91 \\
    B97-3c                      &  0.49             &  0.64 \\
    GFN2-xTB                    &  1.19             &  1.75 \\
    GFN1-xTB                    &  1.35             &  1.87 \\
    GFN0-xTB                    &  2.35             &  3.10 \\
    ANI-2x                      &  1.44             &  1.96 \\
    ANI-1ccx                    &  2.82             &  3.95 \\\hline
    $^{*}$OrbNet Denali         &  0.46             &  0.65 \\
    $^{*}$OrbNet Denali (10\%)  &  0.67             &  0.85 \\\hline
\end{tabular}
\end{table}

\subsection{Torsional profiles of druglike molecules}
\begin{table*}
\caption{\label{tab:torsion_stats}The performance of eight methods on the TorsionNet500 benchmark set. 
The reference energies are calculated at the $\omega$B97X-D3/def2-TZVP level of theory, except for rows marked with an asterisk ($^{*}$) where values are obtained from Rai \textit{et al.}\cite{TorsionNet500}, benchmarked against a B3LYP/6-31G* reference.
For a number of methods, the following statistics are shown: 
the percentage of the 500 torsion profiles for which the Pearson correlation coefficient ($R$) is greater than 0.9, the average Pearson $R$ over the torsion profiles, the MAE and RMSE of the relative energies of the torsion profiles and, lastly, the percentage of torsion profiles where the global minimum of the profile is correct to within $20^{\circ}$ and 1 kcal/mol.
}
\begin{tabular}{lrrrrr}
\hline
    Model   & \shortstack{\% profiles with \\ Pearson $R > 0.9$} & \shortstack{Average Profile \\ Pearson $R$ } &  MAE [kcal/mol] & RMSE [kcal/mol] &  \shortstack{\% local minima with \\ $\Delta \theta <20^{\circ} $and \\ $\Delta E \leq 1$ kcal/mol }\\ \hline
    OrbNet Denali                      &  99.4\% & 0.995 & 0.12  & 0.18  & 100.0\% \\
    OrbNet Denali (10\%)               &  98.8\% & 0.988 & 0.23  & 0.34  & 100.0\% \\
    B97-3c                             &  97.4\% & 0.985 & 0.29  & 0.43  & 100.0\% \\
    GFN2-xTB                           &  76.4\% & 0.881 & 0.73  & 1.0   &  94.0\% \\
    GFN1-xTB                           &  65.6\% & 0.832 & 0.94  & 1.3   &  89.4\% \\
    GFN0-xTB                           &  52.6\% & 0.748 & 1.2   & 1.7   &  85.0\% \\
    MMFF94                             &  54.2\% & 0.760 & 2.4   & 5.2   &  75.2\% \\
    ANI-2x                             &  73.2\% & 0.899 & 1.3   & 1.9   & 91.8\% \\
    $^{*}$ANI-2x\cite{TorsionNet500}         &  54\%   & 0.75  & 1.4   & 2.0   & 66\% \\
    $^{*}$TorsionNet\cite{TorsionNet500}     &  79\%   & 0.91  & 0.7   & 1.3   & 83\% \\\hline
\end{tabular}
\end{table*}

\begin{figure*}
    \centering
    \includegraphics[width=\linewidth]{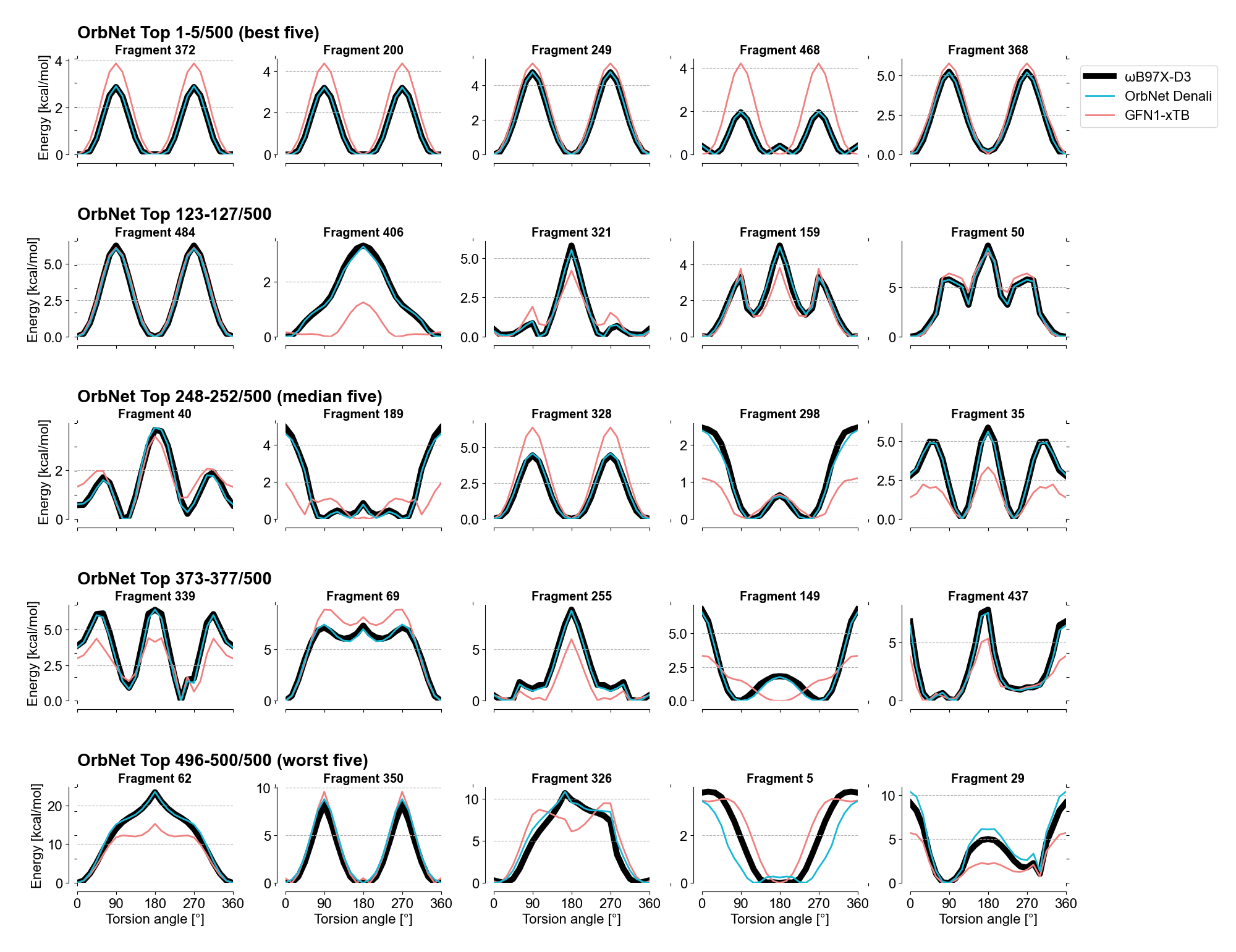}
    \caption{\label{fig:torsionnet500} The charts show torsional profiles of 25 druglike molecules from the TorionNet500 database,\cite{TorsionNet500} stratified to represent quintiles of the OrbNet Denali error relative to the same torsional profiles calculated at the $\omega$B97X-D3/def2-TZVP level of theory, shown as a reference (black).
    Additionally, the same torsion profiles are shown for OrbNet Denali's baseline method, GFN1-xTB (red).
    }
\end{figure*}
Another common benchmark for potential energy surfaces is the accuracy with which torsional profiles can be reproduced.
The recently published TorsionNet500 benchmark\cite{TorsionNet500} compiles torsional profiles of 500 chemically diverse fragments containing the elements H, C, N, O, F, S, and Cl.
For these torsion profiles, we have computed reference energies at the $\omega$B97X-D3/def2-TZVP level, corresponding to the level of theory used to train OrbNet Denali.
Following Rai \textit{et al.}\cite{TorsionNet500}, we benchmark the performance of OrbNet Denali by comparing several different measures of accuracy discussed in the following text.
An overview can be found in Table \ref{tab:torsion_stats}.
The first measure is the number of torsion profiles where the Pearson correlation coefficient ($R$) between the reference energies and the predicted energies is greater than 0.9.
For OrbNet Denali, we find this to be true for 99.4\% of the profiles, while for OrbNet Denali (10\%), the corresponding number is 98.8\%, with average Pearson $R$ values of 0.995 and 0.988, respectively.
Second, the average MAE and RMSE for the torsion profiles are 0.12 and 0.18 kcal/mol for the full OrbNet Denali models, and 0.23 and 0.34 kcal/mol for OrbNet Denali (10\%).
Finally, both OrbNet Denali models correctly predict the location of the global minimum to within 20$^{\circ}$ and its energy to within 1 kcal/mol for all 500 profiles. (These thresholds are taken from Ref. \citenum{TorsionNet500}.)
We note that these excellent results are achieved in spite of the fact that the OrbNet Denali training set contains no torsion profiles.

For OrbNet Denali's baseline method, GFN1-xTB,\cite{gfn1} the same figures are lower with 65.6\% of the profiles having $R > 0.9$, and with an average $R$ value of 0.832, and the average MAE and RMSE 0.94 and 1.3 kcal/mol, while capturing a good minimum for 89.4\% of the predicted profiles.\cite{GFN1xTBGrimme2017}
Fig. \ref{fig:torsionnet500} presents 25 torsion energy profiles of OrbNet Denali versus GFN1-xTB, stratified by the error of OrbNet Denali. 
OrbNet Denali reproduces every point along every torsion within chemical accuracy of 1 kcal/mol in every case. 
Indeed, for all but the 5 worst cases, the OrbNet Denali torsion profile is quantitatively identical to that of the reference method. 
On the other hand, GFN1-xTB shows large errors for the several torsion profiles.
In some cases, the shape of the GFN1-xTB profile is qualitatively incorrect.
Overall, this suggests that the accuracy of OrbNet Denali is not solely due to the use of a good baseline method, but rather that OrbNet Denali is correctly able to capture the subtle differences in the torsion profiles.

We also compare torsion profiles calculated using another well-tested and reliable DFT method, B97-3c, to our reference profiles.\cite{Brandenburg2018}
For B97-3c, the MAE and RMSE w.r.t. the $\omega$B97X-D3 profile are 0.29 and 0.43 kcal/mol. 
While these numbers do not answer which is the more accurate DFT method, they do highlight the fact that OrbNet Denali is almost three times closer to the DFT reference than the variation between these two DFT methods. OrbNet Denali can therefore be considered on-par with DFT methods for this application.

Lastly, we compare OrbNet Denali to the Merck Molecular Mechanics Force Field 94 (MMFF94) and the two ML-based methods, ANI-2x and TorsionNet.\cite{MMFF94,ANI2x,TorsionNet500}
The MMFF94 force field is found to have the lowest accuracy of capturing the $\omega$B97X-D3/def2-TZVP predicted minima, only finding the right minimum within the tolerance 75.2\% of the time, and with higher MAE and RMSE across the torsion profiles, at 1.4 kcal/mol and 5.2 kcal/mol, respectively.
For ANI-2x, we find that a low-energy minimum is captured within the $20^{\circ}$ tolerance with a 91.8\% success rate, compared to the  $\omega$B97X-D3/def2-TZVP reference torsion profiles, which is better than MMFF94, GFN0-xTB and GFN1-xTB.
We note that ANI-2x is parametrized against $\omega$B97X/6-31G(d) reference data, and therefore it is possible that if that combination of DFT functional and basis set had been used to generate the reference curves, ANI-2x would have improved statistics in this test.

ANI-2x --- while having better accuracy at finding the low-energy minima correctly --- comes out with a larger MAE and RMSE than GFN0-xTB and GFN1-xTB, due to underestimation of the rotational barriers.

In addition to these numbers, we highlight the same benchmarks from Ref.~\citenum{TorsionNet500}, which compares ANI-2x and TorsionNet on the same structures, but against B3LYP/6-31G(d) energies.
We preface this by noting, again, that ANI-2x is parametrized against $\omega$B97X/6-31G(d) reference data, while TorsionNet is parametrized against B3LYP/6-31G(d) reference data, so it is possible that our reference data provides a more fair reference for ANI-2x.\cite{ANI2x,TorsionNet500}
Against the B3LYP/6-31G(d) reference, TorsionNet is able to locate the low-energy minima with 83\% success, and ANI-2x 66\% success.
The MAE and RMSE for TorsionNet against its the torsion profile calculated at its own reference level of theory are 0.7 and 1.3 kcal/mol respectively, while the MAE and RMSE for ANI-2x are 1.4 and 2.0 kcal/mol respectively, which is within 0.1 kcal/mol from the same values versus the $\omega$B97X-D3/def2-TZVP reference.

\section{Conclusions}
We have presented a machine learning model, OrbNet Denali, that enables highly efficient energy prediction across broad swaths of chemical space relevant to bio- and organic chemistry applications.

The computational cost of running OrbNet Denali consists of the time spent to run the GFN1-xTB featurization plus additional overhead of running the inference.
In our current CPU-based implementation, the cost of inference is roughly comparable to that of running the GFN1-xTB featurization for molecules of the size presented in this paper.
We note that large speedups on these steps are expected for a GPU-based implementation of OrbNet inference.

The accuracy of OrbNet Denali is comparable to that of modern DFT functionals, such as B97-3c, as demonstrated on several benchmark datasets.
%
%
For the GMTKN55 benchmark set, 37 of the subsets can be evaluated with OrbNet Denali, only limited by the periodic table coverage and ability to handle non-singlet spin states.
For for these 37 subsets, the WTMAD-1 and WTMAD-2 scores were found to be 7.19 and 9.84, respectively, which is on par with many DFT functionals.\cite{GMTKN55}
For non-covalent interactions, OrbNet Denali reaches chemical accuracy for the dissociation curves of the S66x10 datasets.
We find that OrbNet Denali provides conformer scoring on the Hutchison benchmark set with an accuracy of 0.90 $\pm$ 0.02, which is better than some of the common DFT functional tested.
Lastly, we report that OrbNet Denali reproduces the torsional profiles of 500 drug molecules in accordance the reference model ($\omega$B97X-D3/def2-TZVP) to an MAE of only 0.12 kcal/mol, and in all cases was able to identify the correct global minimum within 20$^{\circ}$.

Although our training data does not contain molecular systems similar to many of the systems in the GMTKN55 dataset, dissociation curves, or torsion profiles, OrbNet Denali nevertheless captures the correct physical behavior with very little error when compared to the reference model.
This observation inspires confidence in future development of new OrbNet models.
Natural developments will include training a larger model on more data, spanning even broader swaths of chemical space (and of the periodic table). 
Since the reference model could in many cases be reproduced with an error much smaller than the inherent error of the reference model itself, another potential path of development seems to be retraining the model on data obtained using a higher level methods, such as coupled cluster theory, for example through transfer learning.

We emphasize that even without these further developments, OrbNet Denali provides an exceptionally efficient drop-in replacement for DFT energy predictions for biological and organic chemistry.

\begin{acknowledgments}
Z.Q. acknowledges  graduate research funding from Caltech and partial support from the Amazon-Caltech AI4Science fellowship.
T.F.M. and A.A.  acknowledge partial support from the Caltech DeLogi fund, and A.A. acknowledges support from a Caltech Bren professorship.
The authors gratefully acknowledge NVIDIA, including Abe Stern, Thorsten Knuth, Josh Romero, and Tom Gibbs, for helpful discussions regarding GPU implementations of graph neural networks.
Computational resources were provided by the National Energy Research Scientific Computing Center (NERSC), a DOE Office of Science User Facility supported by the DOE Office of Science under Contract No. DE-AC02-05CH11231.
This research used resources of the Oak Ridge Leadership Computing Facility at the Oak Ridge National Laboratory, which is supported by the Office of Science of the U.S. Department of Energy under Contract No. DE-AC05-00OR22725.
9 of the authors (ASC, SKS, MBO, DGAS, FD, PJB, MW, FRM, TFM) are employees of Entos, Inc. or its affiliates.
\end{acknowledgments}
\section*{Data Availability}
The 2.3 million geometries and energy labels in the OrbNet Denali training set are freely available from FigShare at \url{https://doi.org/10.6084/m9.figshare.14883867}.

\section*{Supplementary Information}
See the Supplementary Information for additional data on GMTKN55 benchmarks, including the MAE for every GMTKN55 subset for every method discussed in the main text, where applicable, and a table containing the GMTKN55 coverage for these methods.

\bibliography{main}
\clearpage

\section*{Supporting Information}

\begin{figure*}
    \centering
    \includegraphics[width=0.8\linewidth]{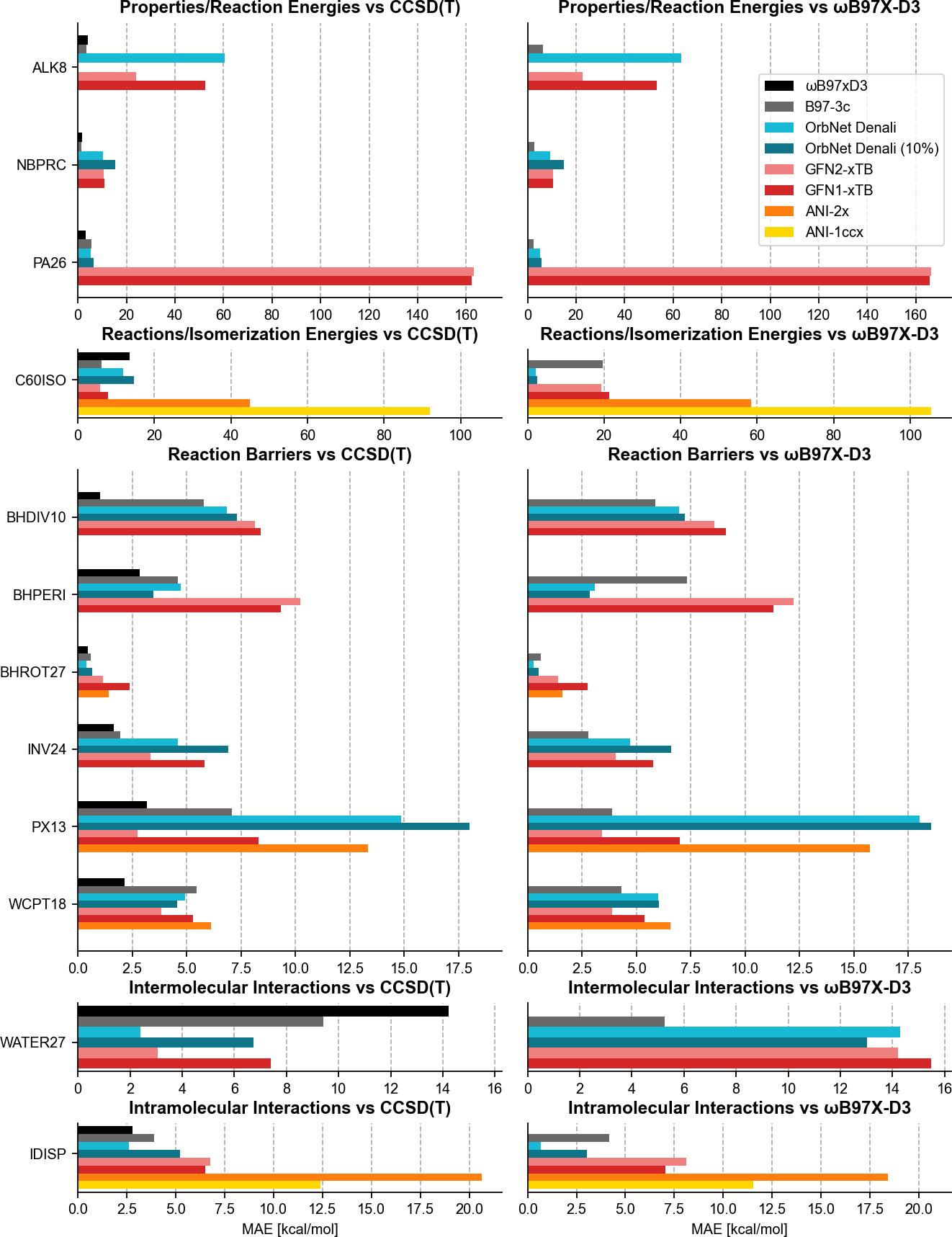}
    \caption{The figure shows the performance of OrbNet Denali on the GMTKN55 subsets that only contain elements, charge- and spin-states that are supported by OrbNet Denali, but contain chemistry that is not represented in the training set (for example water clusters, C$_{60}$ molecules, transition states, etc.). In the left colum, the MAE values are with respect to the CCSD(T) references in the GMTKN55 dataset\cite{GMTKN55} while, in the right column, the MAE values are with respect to $\omega$B97X-D3/def2-TZVP reference energies (the same method used to generate the OrbNet training data). Due to an error in the training of OrbNet Denali 10\%, that model does not support lithium and the ALK8 test set is excluded here. \label{fig:si_gmtkn55_extrapolative}
        }
\end{figure*}

\begin{table*}
\caption{Overview of the mean absolute error (MAE) of every GMTKN55 subset for every method where applicable as discussed in the text. The reference values are the high-level reference values from the GMTKN55 database.\cite{GMTKN55} The MAE values displayed are in units of kcal/mol, along with the WTMAD-$n$ which is calculated as discussed in the main text.\label{tab:si_gmtkn55_ccsdt}
}
\begin{tabular}{lcccccccc}
        & MAE [kcal/mol] &&&&&& \\
Subset	&	$\omega$B97X-D3	&	B97-3c	&	OrbNet Denali	&	ANI-1ccx	&	ANI-2x	&	GFN2-xTB	&	GFN1-xTB	&	GFN0-xTB	\\\hline
W4-11	&	3.52	&	7.48	&		&		&		&		&		&		\\
G21EA	&	7.07	&	8.24	&		&		&		&		&		&		\\
G21IP	&	3.07	&	3.69	&		&		&		&		&		&		\\
DIPCS10	&	5.58	&	4.16	&		&		&		&	274.75	&	301.44	&	1936.47	\\
PA26	&	3.32	&	5.58	&	5.29	&		&		&	163.05	&	162.30	&	938.31	\\
SIE4x4	&	12.18	&	22.54	&		&		&		&		&		&		\\
ALKBDE10	&	5.28	&	7.90	&		&		&		&		&		&		\\
YBDE18	&	2.17	&	5.13	&		&		&		&		&		&		\\
AL2X6	&	3.07	&	2.22	&		&		&		&	14.63	&	15.18	&	8.30	\\
HEAVYSB11	&	2.56	&	2.51	&		&		&		&		&		&		\\
NBPRC	&	1.67	&	1.56	&	10.27	&		&		&	10.51	&	10.96	&	33.00	\\
ALK8	&	4.07	&	3.54	&	60.55	&		&		&	23.91	&	52.55	&	97.75	\\
RC21	&	3.28	&	6.39	&		&		&		&		&		&		\\
G2RC	&	4.58	&	8.34	&		&		&		&	21.92	&	29.27	&	52.60	\\
BH76RC	&	2.31	&	3.64	&		&		&		&		&		&		\\
FH51	&	2.47	&	4.42	&	3.58	&		&	13.38	&	11.41	&	12.05	&	22.14	\\
TAUT15	&	1.05	&	1.71	&	1.13	&		&	2.51	&	0.98	&	5.79	&	3.89	\\
DC13	&	6.81	&	11.33	&		&		&		&		&		&		\\\hline
MB16-43	&	40.75	&	27.39	&		&		&		&		&		&		\\
DARC	&	1.27	&	4.34	&	1.31	&	1.13	&	5.48	&	17.77	&	15.82	&	15.46	\\
RSE43	&	1.44	&	3.49	&		&		&		&		&		&		\\
BSR36	&	4.36	&	1.90	&	0.77	&	2.69	&	8.85	&	2.76	&	2.34	&	2.03	\\
CDIE20	&	0.72	&	1.98	&	0.61	&	2.09	&	3.63	&	1.80	&	2.04	&	2.27	\\
ISO34	&	1.18	&	1.87	&	1.21	&	1.24	&	12.94	&	6.90	&	6.30	&	10.15	\\
ISOL24	&	2.75	&	5.19	&	2.64	&		&		&	11.68	&	10.92	&	10.78	\\
C60ISO	&	1.18	&	6.27	&	11.82	&	91.96	&	44.94	&	5.80	&	7.88	&	11.89	\\
PArel	&	0.67	&	1.80	&	1.60	&		&		&	5.86	&	4.54	&	7.09	\\\hline
BH76	&	2.25	&	6.89	&		&		&		&		&		&		\\
BHPERI	&	2.85	&	4.59	&	4.72	&		&		&	10.24	&	9.32	&	6.84	\\
BHDIV10	&	1.01	&	5.80	&	6.83	&		&		&	8.12	&	8.40	&	9.99	\\
INV24	&	1.63	&	1.96	&	4.59	&		&		&	3.32	&	5.80	&	5.06	\\
BHROT27	&	0.47	&	0.61	&	0.39	&		&	1.42	&	1.17	&	2.38	&	1.68	\\
PX13	&	3.18	&	7.08	&	14.84	&		&	13.32	&	2.74	&	8.30	&	17.16	\\
WCPT18	&	2.14	&	5.46	&	4.91	&		&	6.13	&	3.84	&	5.30	&	8.40	\\\hline
RG18	&	0.11	&	0.12	&		&		&		&	0.11	&	0.32	&	0.44	\\
ADIM6	&	0.36	&	0.53	&	0.40	&	0.64	&	0.34	&	1.15	&	1.01	&	0.51	\\
S22	&	0.36	&	0.29	&	0.45	&	5.00	&	1.51	&	0.76	&	1.33	&	1.63	\\
S66	&	0.52	&	0.32	&	0.48	&	2.52	&	1.11	&	0.73	&	1.07	&	1.29	\\
HEAVY28	&	0.26	&	0.80	&		&		&		&	0.61	&	0.65	&	0.91	\\
WATER27	&	14.23	&	9.41	&	2.39	&		&		&	3.05	&	7.39	&	17.81	\\
CARBHB12	&	0.83	&	2.07	&	0.91	&		&		&	1.79	&	0.67	&	2.32	\\
PNICO23	&	0.38	&	1.64	&	1.71	&		&		&	1.11	&	2.33	&	2.00	\\
HAL59	&	0.34	&	1.62	&	2.15	&		&		&	1.28	&	1.34	&	1.76	\\
AHB21	&	3.40	&	3.27	&	1.81	&		&		&	2.97	&	4.68	&	26.02	\\
CHB6	&	1.32	&	1.37	&	3.58	&		&		&	5.40	&	3.95	&	20.51	\\
IL16	&	2.09	&	2.34	&	4.60	&		&		&	4.32	&	5.69	&	55.34	\\\hline
IDISP	&	2.78	&	3.91	&	2.61	&	12.39	&	20.64	&	6.78	&	6.53	&	13.27	\\
ICONF	&	0.34	&	0.38	&	1.25	&		&		&	1.63	&	2.63	&	1.55	\\
ACONF	&	0.09	&	0.21	&	0.06	&	0.56	&	0.16	&	0.19	&	0.66	&	0.44	\\
Amino20x4	&	0.26	&	0.33	&	0.35	&		&	1.00	&	0.95	&	1.11	&	1.22	\\
PCONF21	&	0.33	&	0.83	&	0.47	&	3.18	&	2.20	&	1.76	&	2.17	&	1.99	\\
MCONF	&	0.48	&	0.33	&	0.42	&	0.92	&	0.82	&	1.72	&	1.44	&	1.34	\\
SCONF	&	0.30	&	0.77	&	0.32	&	1.58	&	2.47	&	1.64	&	2.50	&	2.10	\\
UPU23	&	0.94	&	0.51	&	0.87	&		&		&	2.91	&	1.24	&	3.74	\\
BUT14DIOL	&	0.41	&	0.41	&	0.40	&	0.61	&	1.42	&	1.25	&	0.95	&	0.60	\\\hline
WTMAD-1 [arb. units]	&	3.37	&	5.76	&	7.19	&	14.12	&	13.56	&	12.00	&	14.72	&	24.83	\\
WTMAD-2 [arb. units]	&	5.87	&	10.22	&	9.84	&	23.80	&	24.11	&	18.89	&	22.63	&	37.99	\\\hline
\end{tabular}
\end{table*}

\begin{table*}
\caption{Overview of the mean absolute error (MAE) of every GMTKN55 subset for every method where applicable as discussed in the text. The reference energies are calculated at the $\omega$B97X-D3 level of theory The MAE values displayed are in units of kcal/mol, along with the WTMAD-$n$ which is calculated as discussed in the main text.\label{tab:si_gmtkn55_wb97xd3}
}
\begin{tabular}{lcccccccc}
        & MAE [kcal/mol] &&&&&& \\
SUBSET	&	B97-3c	&	Denali	&	Denali 10\%	&	ANI-1ccx	&	ANI-2x	&	GFN2-xTB	&	GFN1-xTB	&	GFN0-xTB	\\\hline
W4-11	&	7.77	&		&		&		&		&		&		&		\\
G21EA	&	2.16	&		&		&		&		&		&		&		\\
G21IP	&	2.87	&		&		&		&		&		&		&		\\
DIPCS10	&	2.63	&		&		&		&		&	279.84	&	306.53	&	1941.57	\\
PA26	&	2.57	&	5.06	&	5.80	&		&		&	166.38	&	165.62	&	941.63	\\
SIE4x4	&	10.82	&		&		&		&		&		&		&		\\
ALKBDE10	&	4.71	&		&		&		&		&		&		&		\\
YBDE18	&	3.81	&		&		&		&		&		&		&		\\
AL2X6	&	1.55	&		&		&		&		&	11.56	&	13.98	&	8.24	\\
HEAVYSB11	&	2.10	&		&		&		&		&		&		&		\\
NBPRC	&	2.92	&	9.23	&	14.87	&		&		&	10.41	&	10.53	&	33.40	\\
ALK8	&	6.34	&	63.24	&		&		&		&	22.58	&	53.20	&	98.41	\\
RC21	&	4.43	&		&		&		&		&		&		&		\\
G2RC	&	5.75	&		&		&		&		&	19.71	&	25.91	&	53.23	\\
BH76RC	&	2.58	&		&		&		&		&		&		&		\\
FH51	&	4.02	&	1.33	&	4.36	&		&	11.15	&	9.81	&	10.67	&	22.15	\\
TAUT15	&	1.15	&	0.27	&	0.45	&		&	1.84	&	1.26	&	5.25	&	3.54	\\
DC13	&	11.15	&		&		&		&		&		&		&		\\\hline
MB16-43	&	28.97	&		&		&		&		&		&		&		\\
RSE43	&	2.06	&		&		&		&		&		&		&		\\
BSR36	&	2.47	&	3.97	&	4.64	&	1.68	&	4.48	&	6.39	&	4.13	&	5.97	\\
CDIE20	&	1.65	&	0.82	&	1.24	&	2.48	&	4.14	&	1.14	&	1.42	&	2.43	\\
ISO34	&	1.56	&	0.32	&	0.87	&	1.78	&	12.07	&	6.52	&	6.40	&	10.20	\\
ISOL24	&	5.43	&	1.17	&	1.39	&		&		&	11.37	&	10.03	&	10.79	\\
C60ISO	&	19.66	&	2.08	&	2.61	&	105.51	&	58.49	&	19.34	&	21.43	&	25.43	\\
PArel	&	1.59	&	1.58	&	3.20	&		&		&	6.04	&	4.82	&	7.40	\\\hline
BH76	&	5.28	&		&		&		&		&		&		&		\\
BHPERI	&	7.34	&	3.09	&	2.87	&		&		&	12.22	&	11.29	&	9.53	\\
BHDIV10	&	5.86	&	6.95	&	7.24	&		&		&	8.58	&	9.10	&	10.54	\\
INV24	&	2.79	&	4.72	&	6.59	&		&		&	4.05	&	5.78	&	6.06	\\
BHROT27	&	0.60	&	0.29	&	0.51	&		&	1.61	&	1.40	&	2.77	&	2.05	\\
PX13	&	3.90	&	18.02	&	18.56	&		&	15.72	&	3.41	&	6.98	&	17.15	\\
WCPT18	&	4.33	&	6.00	&	6.05	&		&	6.56	&	3.88	&	5.36	&	7.94	\\\hline
RG18	&	0.17	&		&		&		&		&	0.16	&	0.38	&	0.45	\\
ADIM6	&	0.89	&	0.07	&	0.11	&	1.00	&	0.70	&	1.51	&	1.37	&	0.87	\\
S22	&	0.30	&	0.41	&	0.59	&	5.36	&	1.62	&	1.01	&	1.64	&	1.94	\\
S66	&	0.47	&	0.32	&	0.46	&	3.03	&	0.97	&	1.17	&	1.56	&	1.70	\\
HEAVY28	&	0.86	&		&		&		&		&	0.57	&	0.71	&	0.89	\\
WATER27	&	5.26	&	14.28	&	13.02	&		&		&	14.20	&	15.49	&	24.07	\\
CARBHB12	&	1.27	&	0.78	&	1.51	&		&		&	2.45	&	1.05	&	3.06	\\
PNICO23	&	1.55	&	1.84	&	2.15	&		&		&	1.04	&	2.30	&	2.18	\\
HAL59	&	1.69	&	2.09	&	2.45	&		&		&	1.28	&	1.29	&	1.84	\\
AHB21	&	0.97	&	2.60	&	1.96	&		&		&	4.79	&	5.10	&	23.89	\\
CHB6	&	1.01	&	3.45	&		&		&		&	5.44	&	4.41	&	21.31	\\
IL16	&	0.81	&	3.10	&	6.80	&		&		&	6.33	&	7.30	&	53.26	\\\hline
IDISP	&	4.18	&	0.69	&	3.02	&	11.52	&	18.43	&	8.11	&	7.04	&	14.43	\\
ICONF	&	0.49	&	1.08	&	2.73	&		&		&	1.61	&	2.72	&	1.54	\\
ACONF	&	0.26	&	0.07	&	0.11	&	0.56	&	0.16	&	0.18	&	0.62	&	0.41	\\
Amino20x4	&	0.32	&	0.16	&	0.36	&		&	1.09	&	0.93	&	1.15	&	1.25	\\
PCONF21	&	0.92	&	0.29	&	0.86	&	3.27	&	2.26	&	1.63	&	2.04	&	1.78	\\
MCONF	&	0.36	&	0.10	&	0.30	&	1.18	&	0.93	&	1.97	&	1.85	&	1.61	\\
SCONF	&	0.53	&	0.15	&	0.19	&	1.51	&	2.21	&	1.39	&	2.25	&	1.85	\\
UPU23	&	0.54	&	0.62	&	0.84	&		&		&	2.12	&	1.21	&	4.49	\\
BUT14DIOL	&	0.11	&	0.07	&	0.22	&	1.01	&	1.03	&	1.43	&	1.20	&	0.65	\\\hline
WTMAD-1  [arb. units]	&	4.98	&	6.42	&	7.40	&	15.82	&	13.23	&	12.32	&	15.06	&	25.89	\\
WTMAD-2  [arb. units]	&	8.84	&	8.40	&	11.41	&	27.44	&	22.52	&	20.19	&	23.79	&	39.72	\\\hline
\end{tabular}
\end{table*}

\begin{table*}
\caption{Overview of which GMTKN55 subsets are supported by OrbNet Denali, ANI-1ccx, ANI-2x, and the GFN$n$-xTB methods. For OrbNet Denali, the allowed subsets are those that only contain singlet-state molecules with the elements H, Li, B, C, N, O, F, Na, Mg, Si, P, S, Cl, K, Ca, Br, and I. For ANI-1ccx\cite{ANI-1} and ANI-2x\cite{ANI2x} only neutral single-state molecules containing the elements H, C, N, O (for ANI-1ccx) or H, C, N, O, F, Cl, S (for ANI-2x) are allowed. For the GFN$n$-xTB family of methods, only singlet-state molecules are allowed in this list.\label{tab:si_gmtkn55_coverage}
}
\begin{tabular}{llccccc}
Category                                & Subset        & OrbNet Denali &  ANI-1ccx & ANI-2x  & GFN$n$-xTB\\\hline
Basic properties and reaction energies  &     YBDE18    & -          & -          & -      & -      \\
                                        &      W4-11    & -          & -          & -      & -      \\
                                        &     TAUT15    & +          & -          & +      & +      \\
                                        &     SIE4x4    & -          & -          & -      & -      \\
                                        &       RC21    & -          & -          & -      & -      \\
                                        &       PA26    & +          & -          & -      & +      \\
                                        &      NBPRC    & +          & -          & -      & +      \\
                                        &  HEAVYSB11    & -          & -          & -      & -      \\
                                        &       G2RC    & -          & -          & -      & +      \\
                                        &      G21IP    & -          & -          & -      & -      \\
                                        &      G21EA    & -          & -          & -      & -      \\
                                        &       FH51    & +          & -          & +      & +      \\
                                        &    DIPCS10    & -          & -          & -      & +      \\
                                        &       DC13    & -          & -          & -      & -      \\
                                        &     BH76RC    & -          & -          & -      & -      \\
                                        &   ALKBDE10    & -          & -          & -      & -      \\
                                        &       ALK8    & +          & -          & -      & +      \\
                                        &      AL2X6    & -          & -          & -      & +      \\\hline
Reaction and isomerisation energies     &      RSE43    & -          & -          & -      & -      \\
                                        &      PArel    & +          & -          & -      & +      \\
                                        &    MB16-43    & -          & -          & -      & -      \\
                                        &     ISOL24    & +          & -          & -      & +      \\
                                        &      ISO34    & +          & +          & +      & +      \\
                                        &       DARC    & +          & +          & +      & +      \\
                                        &     CDIE20    & +          & +          & +      & +      \\
                                        &     C60ISO    & +          & +          & +      & +      \\
                                        &      BSR36    & +          & +          & +      & +      \\\hline
Reaction barrier heights                &     WCPT18    & +          & -          & +      & +      \\
                                        &       PX13    & +          & -          & +      & +      \\
                                        &      INV24    & +          & -          & -      & +      \\
                                        &    BHROT27    & +          & -          & +      & +      \\
                                        &     BHPERI    & +          & -          & -      & +      \\
                                        &    BHDIV10    & +          & -          & -      & +      \\
                                        &       BH76    & +          & -          & -      & -      \\\hline
Intermolecular noncovalent interactions &    WATER27    & +          & -          & -      & +      \\
                                        &        S66    & +          & +          & +      & +      \\
                                        &        S22    & +          & +          & +      & +      \\
                                        &       RG18    & -          & -          & -      & +      \\
                                        &    PNICO23    & +          & -          & -      & +      \\
                                        &       IL16    & +          & -          & -      & +      \\
                                        &    HEAVY28    & -          & -          & -      & +      \\
                                        &      HAL59    & +          & -          & -      & +      \\
                                        &       CHB6    & +          & -          & -      & +      \\
                                        &   CARBHB12    & +          & -          & -      & +      \\
                                        &      AHB21    & +          & -          & -      & +      \\
                                        &      ADIM6    & +          & +          & +      & +      \\\hline
Intramolecular noncovalent interactions &      UPU23    & +          & -          & -      & +       \\
                                        &      SCONF    & +          & +          & +      & +       \\
                                        &    PCONF21    & +          & +          & +      & +       \\
                                        &      MCONF    & +          & +          & +      & +       \\
                                        &      IDISP    & +          & +          & +      & +       \\
                                        &      ICONF    & +          & -          & -      & +       \\
                                        &  BUT14DIOL    & +          & +          & +      & +       \\
                                        &  Amino20x4    & +          & -          & +      & +       \\
                                        &      ACONF    & +          & +          & +      & +      \\\hline
\hline
\end{tabular}
\end{table*}

\end{document}